\documentstyle[preprint,revtex]{aps}
\begin{document}
\hfill\vbox{\hbox{\bf NUHEP-TH-92-24}\hbox{Nov 1992}}\par
\thispagestyle{empty}
\begin{title}
{\bf $WZ$, $W\gamma$, $WW$ and $ZZ$ pair productions at TeV $e\gamma$
colliders}
\end{title}
\author{Kingman~Cheung}
\begin{instit}
Dept. of Physics \& Astronomy, Northwestern University, Evanston,
Illinois 60208, USA\\
\end{instit}
\begin{abstract}
\nonum
\section{Abstract}
We calculate  the gauge-boson pairs $W^-Z$, $W^-\gamma$, $W^+W^-$,
$ZZ$  productions in the $e^- \gamma$
collisions, where the photon beam is realized by the laser
back-scattering method.
These processes are important tests for the non-abelian gauge sector of the
standard model (SM).   Precise calculations of these processes can therefore
probe the anomalous gauge-boson interactions.
Besides, these processes are important potential backgrounds for the
intermediate mass Higgs (IMH) search in the $e\gamma\rightarrow WH\nu$
production.
\end{abstract}

\newpage
\section{Introduction}
\label{intro}

The electroweak standard model (SM) has so far been very successful
 and consistent
with experiments.  However, it is likely that there exist other models,
of  which  the SM is the effective low energy limit.
The future colliding facilities, which will operate at TeV scale, are
likely to reveal new physics beyond SM.
The symmetry-breaking and the non-abelian gauge-boson sectors are the most
peculiar natures of SM.
A lot have been discussed on the possibilities of the
future hadronic and $e^+e^-$ colliders to probe  the gauge-boson
and  the symmetry-breaking sectors.
With the recent discussion of the physics possibilities at $e\gamma$ and
$\gamma\gamma$ colliders \cite{bord}, they might be as important  as the
hadronic and
$e^+e^-$ colliders.  They  have backgrounds much cleaner than
the hadronic colliders and should be  as clean as the
$e^+e^-$ colliders, and also photon has anomalous gluon and quark contents
\cite{witten} that enable one to study  QCD directly.

The $e^-\gamma$ collisions at  $e^+e^-$
machines can be  realized by directing  a low energy (a few $eV$) laser beam
almost head-to-head to the incident
positron beam.  By Compton scattering, there are abundant, hard
back-scattered photons in the same
direction as the incident positron beam, and  carry a substantial fraction of
the energy of the incident positrons.  Therefore, we have the $e^-\gamma$
collisions.
For details please see Refs.~\cite{teln}. Other possibilities include the
bremsstrahlung and beamstrahlung effects \cite{beams}
but these methods produce photons
mainly in the soft region \cite{teln}, and beamstrahlung depends critically
on the beam structure \cite{beams}.
Therefore we shall limit all the  calculations to $e\gamma$ collisions
produced by the laser back-scattering method.

In recent studies of the Higgs production in $e\gamma$ collisions
\cite{boos,hagi} through
\begin{equation}
e\gamma \rightarrow WH\nu\,,
\end{equation}
the cross section is just a factor of 2 or 3 smaller than that of
$e^+e^-\rightarrow \nu\bar\nu W^*W^* \rightarrow
\nu \bar \nu H$ for $\sqrt{s}=1-2$~TeV,
and so this  production might be a possible channel in  searching for the
Higgs boson.  However, the
backgrounds have not been fully analysed, therefore  we cannot draw any
decisive conclusions.
For the Higgs in the intermediate mass range (IMH) the signature, due to the
dominate decay of $H\rightarrow b\bar b$ and hadronic decay of  $W$, will be
\begin{equation}
e^-\gamma\rightarrow W^-H\nu \rightarrow (jj)(b\bar b) \nu\,,
\end{equation}
where there are 4 jets plus missing energy in the final state. Two of the
four jets  are
reconstructed to the $W$ mass and the other two can be reconstructed
as a resonance peak at the Higgs mass.
For this signature the backgrounds are the $W^-Z$, $W^-W^+$ and $ZZ$
productions
when the $W$ and $Z$ bosons decay hadronically into four jets.
If the Higgs mass falls very
close to the $W$ or $Z$ masses, the signal is much more difficult to identify
and precise calculation of the backgrounds under the $W$ and $Z$ peaks
is necessary.
Therefore, calculations of the $W^-Z$, $W^-W^+$ and $ZZ$ pair productions are
desirable as important potential backgrounds to the IMH search through
$e^-\gamma\rightarrow W^-H\nu$ production.
In addition, it also suffers backgrounds from the $e^-\gamma\rightarrow
\bar t b \nu$ production \cite{jikia} with
$\bar t \rightarrow \bar b W^-$, and $e^-\gamma\rightarrow t\bar t e^-
\rightarrow b\bar b WW e^-$.

Also attention has been focused on the single $W$ production in the channel
$e^-\gamma\rightarrow W^-\nu$ \cite{singleW} to probe the $WW\gamma$ coupling
and search for any  anomalous gauge-boson interactions.
To probe the $WWZ$ coupling, however, we must go for the boson-pair
productions
of $e^-\gamma\rightarrow W^-Z\nu,\,W^+W^-e^-$.  Another interesting point is
that the quartic $WW\gamma\gamma$ and
$WW\gamma Z$ couplings first come in in the $e^-\gamma \rightarrow
W^-Z\nu,\,W^-\gamma\nu$ and $W^-W^+e^-$ productions.  The calculation of these
processes involves
delicate cancellation among the contributions from triple gauge-boson,
quartic gauge-boson and the other Feynman diagrams, which consist of
well-tested vertices.  Therefore, any anomalous interactions of the triple
or quartic gauge-boson vertices
will result in deviations from SM predictions.  It is then favorable to
quantify precisely the production of these gauge-boson pairs, $W^-Z$,
$W^-\gamma$ and $W^-W^+$, within SM so that any deviations from these
predictions will indicate some new physics in the gauge-boson sector.
One advantage of these gauge-boson pair productions in
$e\gamma$ collisions over hadronic collisions is that they  do not have
large QCD background as they  do in
hadronic collisions.  Also these processes as probes to test the triple and
quartic gauge couplings should be as important as  the three gauge-boson
productions  in $e^+e^-$ colliders\cite{han}.

In this paper we calculate the following processes of boson-pair
productions in $e^-\gamma$ collisions,
\begin{eqnarray}
e^-\gamma & \rightarrow &  W^-Z \nu \,, \label{WZ}\\
          & \rightarrow &  W^-\gamma \nu\,, \label{Wg}\\
          & \rightarrow &  W^-W^+ e^-\,, \label{WW}\\
          & \rightarrow &  ZZ e^-\,,\label{ZZ} \\
	  & \rightarrow &  W^-H\nu\,,\label{WH} \\
	  & \rightarrow &  ZHe^-\,.\label{ZH}
\end{eqnarray}
The processes in Eqs.~(\ref{WZ})--(\ref{ZZ}) are important because  they are
the major potential backgrounds to the IMH search in the
channels of Eqs.~(\ref{WH}) and (\ref{ZH}) \cite{boos}.  Besides, the
processes in Eqs.~(\ref{WZ})--(\ref{WW})   are important tests for SM because
they involve non-abelian gauge couplings.  These processes must be
quantified precisely within SM before any anomalous triple or quartic
gauge-boson interactions can be realized in these channels.

The organization of the paper is as follows: we briefly describe the
calculation methods including the photon luminosity function in Sec.~\ref{II},
 following which we present the results in Sec.~\ref{III}, and then
summarize in Sec.~\ref{IV}.  We will also present detail formulas
for the matrix elements of  the processes involved in the appendix~\ref{amp}.

\section{Calculations}
\label{II}

\subsection{Photon Luminosity}

We use the energy spectrum of the back-scattered photon given by \cite{teln}
\begin{equation}
\label{lum}
F_{\gamma /e}(x) = \frac{1}{D(\xi)} \left[ 1-x +\frac{1}{1-x}
-\frac{4x}{\xi(1-x)} + \frac{4x^2}{\xi^2 (1-x)^2} \right] \,,
\end{equation}
where
\begin{equation}
\label{D_xi}
D(\xi) = (1-\frac{4}{\xi} -\frac{8}{\xi^2}) \ln(1+\xi) + \frac{1}{2} +
\frac{8}{\xi} - \frac{1}{2(1+\xi)^2}\,,
\end{equation}
$\xi=4E_0\omega_0/m_e^2$, $\omega_0$ is the energy of the incident laser
photon, $x=\omega/E_0$ is the fraction of the incident positron's
energy carried by the back-scattered photon, and the maximum value $x_{\rm
max}$ is given by
\begin{equation}
x_{\rm max}= \frac{\xi}{1+\xi}\,.
\end{equation}
It is seen from Eq.~(\ref{lum}) and (\ref{D_xi}) that the portion of
photons with  maximum energy  grows with $E_0$ and $\omega_0$.
A large $\omega_0$, however, should be avoided so that the
back-scattered photon does not interact with the incident photon and create
unwanted $e^+e^-$ pairs.  The threshold for  $e^+e^-$ pair creation is
$\omega \omega_0 > m_e^2$, so we require $\omega_{\rm max}
\omega_0 \alt m_e^2$.  Solving $\omega_{\rm max}\omega_0=m_e^2$, we
find
\begin{equation}
\xi = 2(1+\sqrt{2}) \simeq 4.8 \,.
\end{equation}
For the choice $\xi=4.8$ one finds $x_{\rm max}\simeq 0.83$,
$D(\xi)\simeq 1.8$, and
$\omega_0=1.25(0.63)$~eV for a 0.5(1) TeV $e^+e^-$ collider.
Here we have assumed that the electron, positron and back-scattered
photon beams are unpolarized.  We also assume that,
on average, the number of back-scattered photons produced per positron
is 1 (i.e., the conversion coefficient $k$ equals 1).

\subsection{Subprocesses Calculation}

The $W$ and $Z$ bosons are detected through their leptonic or hadronic decays.
So we are not going to impose any acceptance cuts on the $W$ and $Z$ bosons for
their detections, instead, we assume some detection efficiencies for their
decay products to estimate the number of observed events.  On the other hand,
$\gamma$ can be observed directly in the final state by imposing a typical
experimental acceptance, say,
\begin{equation}
\label{photoncut}
\begin{array}{rcl}
p_T(\gamma) & > & 15\,{\rm GeV}\,, \\
|\eta(\gamma)| & < & 2 \,.
\end{array}
\end{equation}
We use the helicity amplitude method of Ref.~\cite{stange} to evaluate the
Feynman amplitudes, and  keep the electron mass $m_e$ finite in all the
calculations.  There are totally 11 contributing Feynman diagrams in the
process $e^-\gamma\rightarrow W^-Z\nu$, 9 in $e^-\gamma\rightarrow
W^-\gamma\nu$, 18 in $e^-\gamma \rightarrow W^-W^+e^-$, and 6 in
$e^-\gamma\rightarrow ZZe^-$,  in the general $R_\xi$ gauge.   The helicity
amplitudes for the processes of Eqs.~(\ref{WZ})--(\ref{ZZ}) are given
in Appendix~\ref{amp}.  The processes of Eqs.~(\ref{WH}) and (\ref{ZH}) have
been calculated in detail in Refs.~\cite{boos,hagi}.  The
total cross-section $\sigma$ is obtained by folding the subprocess
cross-section $\hat \sigma$ in  with the photon luminosity function
of Eqs.~(\ref{lum}) and (\ref{D_xi}); shown in appendix~\ref{amp}.

\section{Results and Discussion}
\label{III}

We show the dependence of the cross sections for all the processes
of Eqs.~(\ref{WZ})--(\ref{ZH}), together with $e^+e^-\rightarrow \nu\bar\nu
W^*W^*
\rightarrow \nu\bar\nu H$, in Fig.~\ref{cross}.  We typically
choose $m_H$ =100~GeV in
the intermediate mass range, and impose the acceptance cuts of
Eq.~(\ref{photoncut}) on the photon that occurs in the final state.
The cross section of
$W^-W^+$ is overwhelming due to a hugh contribution from  the Feynman diagrams
with an almost on-shell $t$-channel $\gamma$ propagator.
This hugh cross section, of order 10~pb in the energy range of 0.5--2~TeV,
is an advantage to probe  the triple or quartic gauge-boson couplings.
The $WZ$ and $W\gamma$ cross section is of order 0.5 and 1~pb in the same
energy range, respectively.
For a yearly
luminosity of 10~${\rm fb}^{-1}$, we have about 5000 $WZ$ and 10000 $W\gamma$
events.  The hadronic branching fraction of both $W$ and $Z$ is $\sim 0.7$
and assuming 50\%
hadronic detection efficiency, we still have 625 observed  events for
$WZ$ and 3500 events for $W\gamma$, which are numerous enough to observe any
anomalous gauge-boson interactions.
Therefore, even with 1\% anomalous gauge coupling it could result in
about 6 and 35
events in $WZ$ and $W\gamma$ productions respectively, and in the order of
hundreds of events for $WW$ production.
Therefore, as mentioned above these gauge-boson pair productions as probes to
test the triple and quartic gauge couplings are as important as the three
gauge-boson productions in $e^+e^-$ collisions, which are of order  0.1~pb
for $\sqrt{s_{e^+e^-}}=  0.5$--2~TeV \cite{han}.
$ZZ$ production is insignificant at
all for  the  energy range that we are considering.
The $WH$ production is of order 0.1--0.2~pb for
$\sqrt{s}=1-2$~TeV and $m_H=100$~GeV,
and $ZH$ production is so much smaller that it will never be discovered.
In comparison we also show the cross section of $e^+e^-\rightarrow \nu\bar\nu
W^*W^*\rightarrow \nu \bar\nu H$, which is dominant over
the $e^+e^-\rightarrow ZH$ production for $\sqrt{s}>0.5$~TeV.
We can see that at $\sqrt{s}=1(2)$~TeV the
$e^-\gamma \rightarrow W^-H\nu$ cross section is only about a factor of 2.5 (2)
smaller than that of $e^+e^-\rightarrow \nu\bar\nu W^*W^* \rightarrow
\nu \bar \nu H$.

For the IMH search in $WH$ production total background from $WW$, $WZ$ and
$ZZ$ is about two order of magnitudes larger (see Fig.~\ref{cross}).
But from Fig.~\ref{ptvv} we can see that  the hugh  cross section of $WW$
can go down sharply by
requiring a moderate  transverse momentum $p_T(VV)$ cut, say
$p_T(VV)>25(50)$~GeV at $\sqrt{s}=0.5(2)$~TeV,  on the $WW$ system to keep
the $\gamma$-propagator far from being on-shell.  Further reduction of the
$WW$ cross section can be achieved by central electron vetoing method, i.e.,
rejecting events with electrons detected in the central rapidity region
($|\eta|<3$).
Then the total background from $WZ$, $ZZ$ and $WW$ is only a few times
larger than the IMH signal in  $e^-\gamma\rightarrow W^-H\nu$
production.
Furthermore, if $b$-tagging has a high efficiency and the invariant mass
reconstruction has a good resolution  these backgrounds
can be substantially reduced, so $WH$ production remains a possible channel to
search for the IMH.  However, a more detail analysis taking  into account
the other  backgrounds from $e^-\gamma\rightarrow \bar tb\nu,\, t\bar t e^-$
and detector resolutions is necessary to establish the Higgs-boson signal.

In Fig.~\ref{mvv}, we show the dependence of the differential cross section
$d\sigma/dM(VV)$ on  the invariant mass $M(VV)$ of the boson pair at
$\sqrt{s}=0.5$ and 2~TeV.  As expected, these curves rise a little bit above
their corresponding $M(VV)$ threshold  and
then fall gradually as $M(VV)$ increases further.  However, the presence of
any anomalous  triple or quartic gauge-boson interactions can alter the $WZ$,
$W\gamma$, $WW$ and $WH$ curves to some extent.  These spectra  can
therefore serve as SM predictions to probe  the anomalous gauge-boson
sector.

\section{Conclusions}
\label{IV}

We have quantified  the productions of
$e^-\gamma \rightarrow W^-Z\nu$, $W^-\gamma\nu$, $W^-W^+e^-$, $ZZe^-$
within SM, and presented the helicity amplitudes for these processes.
These processes  can probe the non-abelian gauge sector of the
SM, and should be as good as the three gauge-boson pair productions in
$e^+e^-$ collisions and better than those in hadronic collisions.
The production rate of $W^-W^+$ pair  is hugh, and that of $W^-Z$ and
$W^-\gamma$ are large enough that a percent-level
anomalous gauge-boson interactions can be detected if they exist.  On the
other hand, the IMH search in the $e\gamma\rightarrow WH\nu$ channel seems
impossible due to hugh background from $WW$ and $WZ$.  However, we have shown
in Fig.~\ref{ptvv} that a $p_T(VV)$ cut can substantially reduce the $WW$
background, together  with central electron vetoing method and
$b$-tagging the total
background from boson-pair productions is only a few times larger than the
IMH signal.  Nevertheless, a more detail signal-background
analysis is needed.

\acknowledgements
This work was supported by the U.~S. Department of Energy, Division of
High Energy Physics, under Grant DE-FG02-91-ER40684.
\newpage
\appendix{}
\label{amp}

In this appendix we present the matrix elements for
processes $e^-\gamma\rightarrow W^-Z\nu,\,W^-\gamma\nu,\, W^-W^+e^-,\, ZZe^-$,
{} from which explicit helicity amplitudes can be directly computed.
To start with, we introduce some general notation:
\begin{eqnarray}
g_a^W(f) & = & -g_b^W(f) = \frac{g}{2 \sqrt{2}} \, , \\
g_a^Z(f) & = & g_Z \left( {T_{3f}\over2} - Q_f x_{\rm w}\right) \, ,\\
g_b^Z(f) & = & - g_Z {T_{3f}\over2} \, ,\\
g_a^\gamma(f) & = & e Q_f\, ,\\
g_b^\gamma(f) & = & 0\, ,\\
g^V(f) & = & g_a^V(f) + g_b^V(f) \gamma^5\,\qquad(V=\gamma,W,Z)\,,\\
D^X(k) & = & \frac{1}{k^2-M_X^2 + i \Gamma_X(k^2) m_X}\,,\qquad
\Gamma_X(k^2) = \Gamma_X \theta(k^2) \nonumber \\
&&  \qquad (\mbox{with }X=\gamma,W,Z,H) \, ,\\
P_V^{\alpha \beta}(k) & = &  \left [ g^{\alpha \beta} + \frac{(1-\xi)k^\alpha
k^\beta}{\xi k^2 - m_V^2} \right ] D^V(k) \,, \\
\Gamma^\alpha (k_1,k_2;\epsilon_1,\epsilon_2) & = & (k_1-k_2)^\alpha \epsilon_1
\cdot \epsilon_2 + (2k_2+k_1) \cdot \epsilon_1 \epsilon_2^\alpha
- (2k_1+k_2) \cdot \epsilon_2 \epsilon_1^\alpha\, , \\
g_{VWW} & = & \left \{
               \begin{array}{ll}
                e \cot \theta_{\rm w}  \quad & {\rm for\ } V=Z \\
                e                      & {\rm for\ } V=\gamma \, .
               \end{array} \right.  \\ \nonumber
\end{eqnarray}
Here $Q_f$ and $T_{3f}$ are the electric charge (in units of the positron
charge) and the third component of weak isospin of the fermion $f$, $g$ is
the SU(2) gauge coupling, and $g_Z=g/\cos \theta_{\rm w}$,
$x_{\rm w}=\sin^2 \theta_{\rm w}$, with $\theta_{\rm w}$ being the weak
mixing angle in the Standard Model.  Dots between 4-vectors denote scalar
products and $g_{\alpha \beta}$ is the Minkowskian metric tensor with
$g_{00}=-g_{11}=-g_{22}=-g_{33}=1$; $\xi$ is a gauge-fixing parameter.

In Figs.~\ref{fey-wz} and \ref{fey-ww}, the momentum-labels $p_i$ denote
the momenta flowing along
the corresponding fermion lines in the direction of the arrows.
We shall always denote the associated spinors by the symbols $u(p_i)$ and $\bar
u(p_i)$ for the ingoing and outgoing arrows, which is usual for the
annihilation and creation of fermions, respectively.

\subsection{$e^-\gamma\rightarrow W^-Z \nu$}

The contributing Feynman diagrams for $e^- (p_1) \gamma (p_2)
\rightarrow W^-(k_1) Z(k_2) \nu (q_1)$ are given in Fig.~\ref{fey-wz}.
We define a shorthand notation
\begin{equation}
\begin{array}{rcl}
J^\mu_1 &=& \bar u(q_1) \gamma^\mu g^W(e) u(p_1) \times D^W(p_1-q_1)\,,
\end{array}
\end{equation}
then the  helicity amplitudes are given by
\begin{eqnarray}
{\cal M}^{(a)} &=& -\, g_{ZWW} g_{\gamma WW} \,
\Gamma_\alpha(-k_1,\,p_2;\, \epsilon(k_1), \, \epsilon(p_2) ) \,
P^{\alpha\beta}_W(p_2-k_1) \nonumber \\
&& \qquad \times \, \Gamma_\beta(-k_2,\,p_1-q_1;\, \epsilon(k_2), \, J_1 )\,,\\
{\cal M}^{(b)} &=& -\, g_{ZWW} g_{\gamma WW} \,
\Gamma_\alpha( k_2,\,k_1;\, \epsilon(k_2), \, \epsilon(k_1) ) \,
P^{\alpha\beta}_W(k_1+k_2) \nonumber  \\
&& \qquad \times \,\Gamma_\beta( p_2,\,p_1-q_1;\, \epsilon(p_2), \, J_1 ) \,,\\
{\cal M}^{(c)} &=& g_{ZWW} g_{\gamma WW} \left[
  2 \epsilon(p_2) \cdot \epsilon(k_2) \epsilon(k_1)\cdot J_1
- \epsilon(p_2) \cdot J_1  \epsilon(k_1)\cdot \epsilon(k_2)\right. \nonumber \\
&& \left. \qquad \qquad \qquad - \,  \epsilon(p_2) \cdot \epsilon(k_1)
              \epsilon(k_2)\cdot J_1          \right] \,, \\
{\cal M}^{(d,e)} &=& g_{\gamma WW} \Gamma_\alpha(-k_1,\,p_2;\,\epsilon(k_1),\,
\epsilon(p_2) ) P^{\alpha\beta}_W(p_2- k_1)  \nonumber \\
&& \times \left [ \bar u(q_1) \gamma_\beta g^W (e) \frac{\overlay{/}{p}_1 -
\overlay{/}{k}_2 + m_e }{(p_1-k_2)^2 - m^2_e}\, \overlay{/}{\epsilon}(k_2)
g^Z(e) u(p_1)  \right. \nonumber \\
&& \qquad \qquad \left. + \; \bar u(q_1) \overlay{/}{\epsilon}(k_2) g^Z(\nu)
\frac{\overlay{/}{q}_1 + \overlay{/}{k}_2}{(q_1+k_2)^2}\, \gamma_\beta g^W(e)
u(p_1) \right ]\,, \\
{\cal M}^{(f)} &=& g_{ZWW} \Gamma_\alpha( k_2,\,k_1;\,\epsilon(k_2),\,
\epsilon(k_1)) P^{\alpha\beta}_W(k_1+k_2) \nonumber \\
&& \times\; \bar u(q_1) \gamma_\beta g^W(e) \frac{\overlay{/}{p}_1 +
\overlay{/}{p}_2 + m_e}{(p_1+p_2)^2 - m_e^2}\, \overlay{/}{\epsilon}(p_2)
g^\gamma(e) u(p_1)\,, \\
{\cal M}^{(g)} &=& - \bar u(q_1) \overlay{/}{\epsilon}(k_1) g^W(e)
\frac{ \overlay{/}{q}_1 + \overlay{/}{k}_1 + m_e}{(q_1+k_1)^2 - m_e^2}
\overlay{/}{\epsilon}(k_2) g^Z(e)
\frac{ \overlay{/}{p}_1 + \overlay{/}{p}_2 + m_e}{(p_1+p_2)^2 -m_e^2}
\nonumber \\
&& \qquad \overlay{/}{\epsilon}(p_2) g^\gamma(e) u(p_1) \,, \\
\noalign{\break}
{\cal M}^{(h)} &=& - \bar u(q_1) \overlay{/}{\epsilon}(k_1) g^W(e)
\frac{ \overlay{/}{q}_1 + \overlay{/}{k}_1 + m_e}{(q_1+k_1)^2 - m_e^2}
\overlay{/}{\epsilon}(p_2) g^\gamma(e)
\frac{ \overlay{/}{p}_1 - \overlay{/}{k}_2 + m_e}{(p_1-k_2)^2 -m_e^2}
\nonumber \\
&& \qquad \overlay{/}{\epsilon}(k_2) g^Z(e) u(p_1) \,, \\
{\cal M}^{(i)} &=& - \bar u(q_1) \overlay{/}{\epsilon}(k_2) g^Z(\nu)
\frac{ \overlay{/}{q}_1 + \overlay{/}{k}_2}{(q_1+k_2)^2}
\overlay{/}{\epsilon}(k_1) g^W(e)
\frac{ \overlay{/}{p}_1 + \overlay{/}{p}_2 + m_e}{(p_1+p_2)^2 -m_e^2}
\nonumber \\
&& \qquad \overlay{/}{\epsilon}(p_2) g^\gamma(e) u(p_1) \,, \\
{\cal M}^{(j)} &=& - g^2 m_W^2 x_{\rm w} \tan\theta_{\rm w}
\frac{\xi}{\xi(p_2-k_1)^2 -m_W^2}\, \epsilon(k_1) \cdot \epsilon(p_2) \,
               \epsilon(k_2) \cdot J_1 \,\\
{\cal M}^{(k)} &=& - g^2 m_W^2 x_{\rm w} \tan\theta_{\rm w}
\frac{\xi}{\xi(k_1+k_2)^2 -m_W^2} \, \epsilon(k_1) \cdot \epsilon(k_2) \,
               \epsilon(p_2) \cdot J_1 \,.
\end{eqnarray}

The contributing Feynman diagrams for $e^-\gamma \rightarrow W^-\gamma\nu$ are
obtained from those in Fig.~\ref{fey-wz} by replacing the $Z$ by $\gamma$.
The helicity amplitudes for $e^-(p_1) \gamma(p_2)\rightarrow W^-(k_1) \gamma(
k_2) \nu (q_1) $ can be obtained from the above expressions by replacing the
corresponding $g_{ZWW}$ and $g^Z(e\; {\rm or}\; \nu)$ by $g_{\gamma WW}$ and
$g^\gamma(e\; {\rm or}\; \nu)$ respectively, and substituting the
$\tan\theta_{\rm w}$ in ${\cal M}^{(j)}$ and ${\cal M}^{(k)}$ by -1.
Since $g^\gamma(\nu)=0$, diagrams~\ref{fey-wz}(e) and (i) do not contribute to
the $W^-\gamma$ production.

\subsection{$e^-\gamma \rightarrow W^- W^+ e^-$}

The contributing Feynman diagrams for the process
$e^-(p_1)\gamma(p_2) \rightarrow W^-(k_1) W^+(k_2) e^-(q_1)$ are shown
in Fig.~\ref{fey-ww}.  We can also  define a shorthand notation
\begin{equation}
\begin{array}{rcl}
J^\mu_V &=& \bar u(q_1) \gamma^\mu g^V(e) u(p_1) \times D^V(p_1-q_1)\,,
\quad {\rm where}\; V=\gamma,\,Z
\end{array}
\end{equation}
then the  helicity amplitudes are given by
\begin{eqnarray}
{\cal M}^{(a)} &=& \sum_{V=\gamma,Z} -\,g_{VWW}\, g_{\gamma WW} \,
P^{\alpha\beta}_W(p_2-k_2) \nonumber \\
&& \quad \times \,\Gamma_\alpha(-k_1,\, p_1-q_1;\, \epsilon(k_1),\, J_V)
       \, \Gamma_\beta( p_2,\, -k_2;\, \epsilon(p_2),\, \epsilon(k_2) ) \,, \\
%
%
{\cal M}^{(b)} &=& \sum_{V=\gamma,Z} - \,g_{VWW} \, g_{\gamma WW} \,
P^{\alpha\beta}_W(p_2-k_1) \nonumber \\
&& \quad \times \, \Gamma_\alpha(p_1-q_1,\, -k_2;\, J_V,\, \epsilon(k_2) )
       \, \Gamma_\beta( -k_1,\, p_2;\, \epsilon(k_1),\, \epsilon(p_2) ) \,, \\
{\cal M}^{(c)} &=& \sum_{V=\gamma,Z} g_{VWW} g_{\gamma WW} \,\left [
2 \epsilon(k_1) \cdot \epsilon(k_2)\, \epsilon(p_2) \cdot J_V \nonumber
\right. \nonumber \\
&& \left. \qquad\qquad -\,\epsilon(k_1) \cdot J_V \, \epsilon(k_2)\cdot
          \epsilon(p_2)
          -\, \epsilon(k_2) \cdot J_V \, \epsilon(k_1)\cdot \epsilon(p_2)\,
    \right ] \,, \\
{\cal M}^{(d)} &=& - \bar u(q_1) \overlay{/}{\epsilon}(k_2) g^W(e)
\frac{\overlay{/}{q}_1 + \overlay{/}{k}_2}{(q_1+k_2)^2}
\overlay{/}{\epsilon}(k_1) g^W(e)
\frac{\overlay{/}{p}_1 + \overlay{/}{p}_2 + m_e}{(p_1+p_2)^2 - m_e^2}
\nonumber \\
&& \qquad \overlay{/}{\epsilon}(p_2) g^\gamma(e) u(p_1) \,, \\
{\cal M}^{(e)} &=& - \bar u(q_1) \overlay{/}{\epsilon}(p_2) g^\gamma(e)
\frac{\overlay{/}{q}_1 - \overlay{/}{p}_2 + m_e}{(q_1-p_2)^2 - m_e^2}
\overlay{/}{\epsilon}(k_2) g^W(e)
\frac{\overlay{/}{p}_1 - \overlay{/}{k}_1 }{(p_1 - k_1)^2} \nonumber \\
&& \qquad \overlay{/}{\epsilon}(k_1) g^W(e) u(p_1) \,, \\
{\cal M}^{(f)} &=& \sum_{V=\gamma,Z} g_{VWW} D^V(k_1+k_2) \,
\Gamma_\alpha (k_1,\,k_2;\, \epsilon(k_1),\, \epsilon(k_2) ) \nonumber \\
&& \quad  \times \,\bar u(q_1) \gamma^\alpha g^V(e)
\frac{\overlay{/}{p}_1 + \overlay{/}{p}_2 + m_e }{(p_1 + p_2)^2 - m_e^2} \,
\overlay{/}{\epsilon}(p_2) g^\gamma(e) u(p_1) \,, \\
{\cal M}^{(g)} &=& \sum_{V=\gamma,Z} g_{VWW} D^V(k_1+k_2) \,
\Gamma_\alpha (k_1,\,k_2;\, \epsilon(k_1),\, \epsilon(k_2) ) \nonumber \\
&& \quad \times \, \bar u(q_1) \overlay{/}{\epsilon}(p_2) g^\gamma(e)
\frac{\overlay{/}{q}_1 - \overlay{/}{p}_2 + m_e }{(q_1 - p_2)^2 - m_e^2} \,
\gamma^\alpha  g^V(e)  u(p_1) \,, \\
{\cal M}^{(h)} &=& g_{\gamma WW} P^{\alpha\beta}_W(p_2-k_2)
\Gamma_\alpha ( p_2,\, -k_2;\, \epsilon(p_2),\, \epsilon(k_2) ) \nonumber \\
&& \quad \times \, \bar u(q_1) \gamma_\beta g^W(e)
\frac{\overlay{/}{p}_1 - \overlay{/}{k}_1}{(p_1 - k_1 )^2} \,
\overlay{/}{\epsilon}(k_1) g^W(e) u(p_1) \,, \\
{\cal M}^{(i)} &=& g_{\gamma WW} P^{\alpha\beta}_W(p_2-k_1) \,
\Gamma_\alpha ( -k_1,\, p_2;\, \epsilon(k_1),\, \epsilon(p_2) ) \nonumber \\
&& \quad \times \, \bar u(q_1)   \overlay{/}{\epsilon}(k_2) g^W(e)
\frac{\overlay{/}{q}_1 + \overlay{/}{k}_2}{(q_1 + k_2 )^2} \,
\gamma_\beta  g^W(e) u(p_1) \,, \\
{\cal M}^{(j)} &=& \sum_{V=\gamma,Z} g^2 m_W^2 x_{\rm w} \,
\frac{\xi}{\xi(p_2-k_2)^2 - m_W^2}\, \epsilon(p_2) \cdot \epsilon(k_2) \,
\epsilon(k_1) \cdot J_V \nonumber \\
&& \qquad \times \, \left \{ \begin{array}{ll}
                              -\tan\theta_{\rm w} & \quad {\rm for}\; V=Z \\
                              1      & \quad {\rm for}\; V=\gamma
                            \end{array}
                   \right.
    \,, \\
{\cal M}^{(k)} &=& \sum_{V=\gamma,Z} g^2 m_W^2 x_{\rm w} \,
\frac{\xi}{\xi(p_2-k_1)^2 - m_W^2}\, \epsilon(p_2) \cdot \epsilon(k_1) \,
\epsilon(k_2) \cdot J_V \nonumber \\
&& \qquad \times \, \left \{ \begin{array}{ll}
                              -\tan\theta_{\rm w} & \quad {\rm for}\; V=Z \\
                              1      & \quad {\rm for}\; V=\gamma
                            \end{array}
                   \right.
    \,.
\end{eqnarray}

\subsection{$e^-\gamma\rightarrow ZZ e^-$}

The contributing Feynman diagrams for the process $e^-(p_1)\gamma(p_2)
 \rightarrow Z(k_1) Z(k_2) e^-(q_1)$ are the same as in
Fig.~\ref{fey-ww}(d) with the $W$-bosons replaced by $Z$-bosons
plus  all possible permutations.  Totally it has six
contributing Feynman diagrams.  They are given by
\begin{eqnarray}
{\cal M}^{(a)} &=& - \bar u(q_1) \overlay{/}{\epsilon}(k_1) g^Z(e)
\frac{\overlay{/}{q}_1 + \overlay{/}{k}_1 +m_e }{(q_1+k_1)^2 - m_e^2 }
\overlay{/}{\epsilon}(k_2) g^Z(e)
\frac{\overlay{/}{p}_1 + \overlay{/}{p}_2 +m_e }{(p_1 + p_2)^2 - m_e^2 }
\nonumber \\
&& \qquad \overlay{/}{\epsilon}(p_2) g^\gamma(e) u(p_1) \,, \\
{\cal M}^{(b)} &=& - \bar u(q_1) \overlay{/}{\epsilon}(k_1) g^Z(e)
\frac{\overlay{/}{q}_1 + \overlay{/}{k}_1 +m_e }{(q_1+k_1)^2 - m_e^2 }
\overlay{/}{\epsilon}(p_2) g^\gamma(e)
\frac{\overlay{/}{p}_1 - \overlay{/}{k}_2 +m_e }{(p_1 - k_2)^2 - m_e^2 }
\nonumber \\
&& \qquad \overlay{/}{\epsilon}(k_2) g^Z(e) u(p_1) \,, \\
{\cal M}^{(c)} &=& - \bar u(q_1) \overlay{/}{\epsilon}(p_2) g^\gamma(e)
\frac{\overlay{/}{q}_1 - \overlay{/}{p}_2 +m_e }{(q_1-p_2)^2 - m_e^2 }
\overlay{/}{\epsilon}(k_1) g^Z(e)
\frac{\overlay{/}{p}_1 - \overlay{/}{k}_2 +m_e }{(p_1 - k_2)^2 - m_e^2 }
\nonumber \\
&& \qquad \overlay{/}{\epsilon}(k_2) g^Z(e) u(p_1) \,,
\end{eqnarray}
plus those terms with $(k_1 \leftrightarrow k_2)$.

These matrix elements are to be summed over polarizations and spins of the
final state gauge-bosons and fermions respectively, and average over the
polarizations of the incoming photon and spins of the initial state electron.
Then the cross section $\sigma$ is obtained by folding the subprocess
cross-section $\hat \sigma$ in with the photon luminosity function as
\begin{equation}
\sigma(s) = \int^{x_{max}}_{M_{\rm final}/s} dx F_{\gamma/e}(x)
\; \hat \sigma( \hat s=xs)  \,,
\end{equation}
where
\begin{equation}
\begin{array}{rcl}
\hat\sigma(\hat s) &=&  \frac{1}{2 (\hat s -m_e^2)}   \int
                     \frac{d^3 k_1}{(2\pi)^3 k_1^0} \,
                     \frac{d^3 k_2}{(2\pi)^3 k_2^0}\,
                     \frac{d^3 q_1}{(2\pi)^3 q_1^0}  \\
 && \qquad \quad \times \;    (2\pi)^4 \delta^{(4)} (p_1+p_2-k_1-k_2-q_1) \,
               \sum | {\cal M}|^2 \, \\
\end{array}
\end{equation}
and $M_{\rm final}$ is the sum of the masses of the final state particles.


%
\newpage
\figure{\label{cross}
The production cross sections for the processes in
Eqs.~(\ref{WZ})--(\ref{ZH}), and $e^+e^-\rightarrow \nu\bar\nu W^*W^*
\rightarrow \nu\bar\nu H$ for
$m_H=100$~GeV versus $\sqrt{s}$ of the parent $e^+e^-$ collider.
The acceptance cuts on the final state photon are $p_T(\gamma)>15$~GeV and
$|\eta(\gamma)|<2$.}

\figure{\label{ptvv}
The differential cross sections $d\sigma/dp_T(VV)$ for the processes in
Eqs.~(\ref{WZ})--(\ref{ZH}) for $m_H=100$~GeV versus the transverse momentum
$p_T(VV)$ of the boson pair at $\sqrt{s}$= (a) 0.5 and (b) 2~TeV.
The acceptance cuts on the final state photon are $p_T(\gamma)>15$~GeV and
$|\eta(\gamma)|<2$.}

\figure{\label{mvv}
The differential cross sections $d\sigma/dM(VV)$ for the processes in
Eqs.~(\ref{WZ})--(\ref{ZH}) for $m_H=100$~GeV versus the invariant mass
$M(VV)$ of the boson pair at $\sqrt{s}$= (a) 0.5 and (b) 2~TeV.
The acceptance cuts on the final state photon are $p_T(\gamma)>15$~GeV and
$|\eta(\gamma)|<2$.}

\figure{\label{fey-wz}
Contributing Feynman diagrams for the process $e^-\gamma \rightarrow
W^-Z \nu$.  Those for $e^-\gamma\rightarrow W^-\gamma\nu$ are the same with
$(Z \leftrightarrow \gamma)$, except
it does not have contributions from (e) and (i).}

\figure{\label{fey-ww}
Contributing Feynman diagrams for the process $e^-\gamma \rightarrow
W^-W^+ e^-$.  Those for $e^-\gamma\rightarrow ZZe^-$  are as in (d) plus five
other permutations}

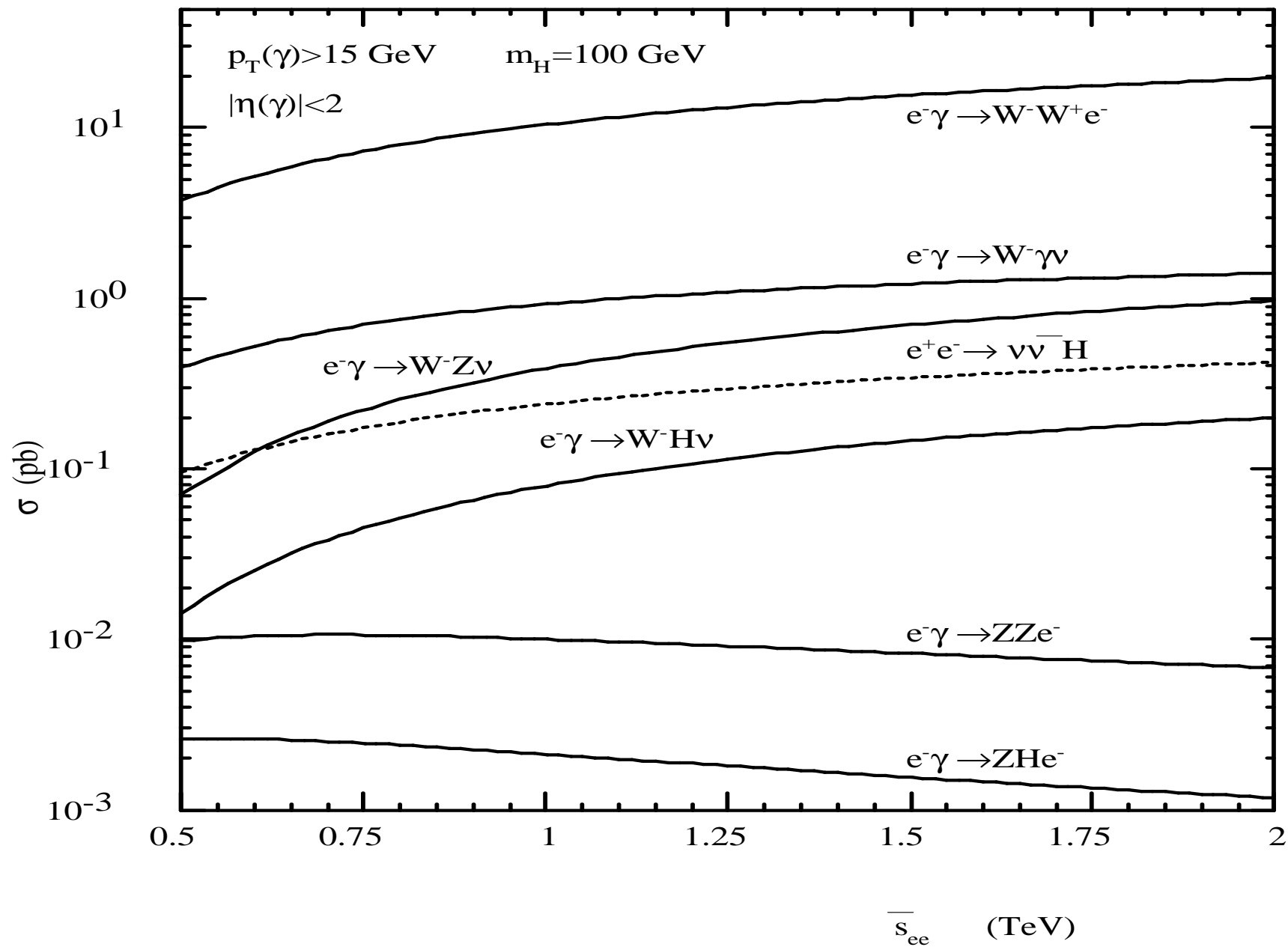
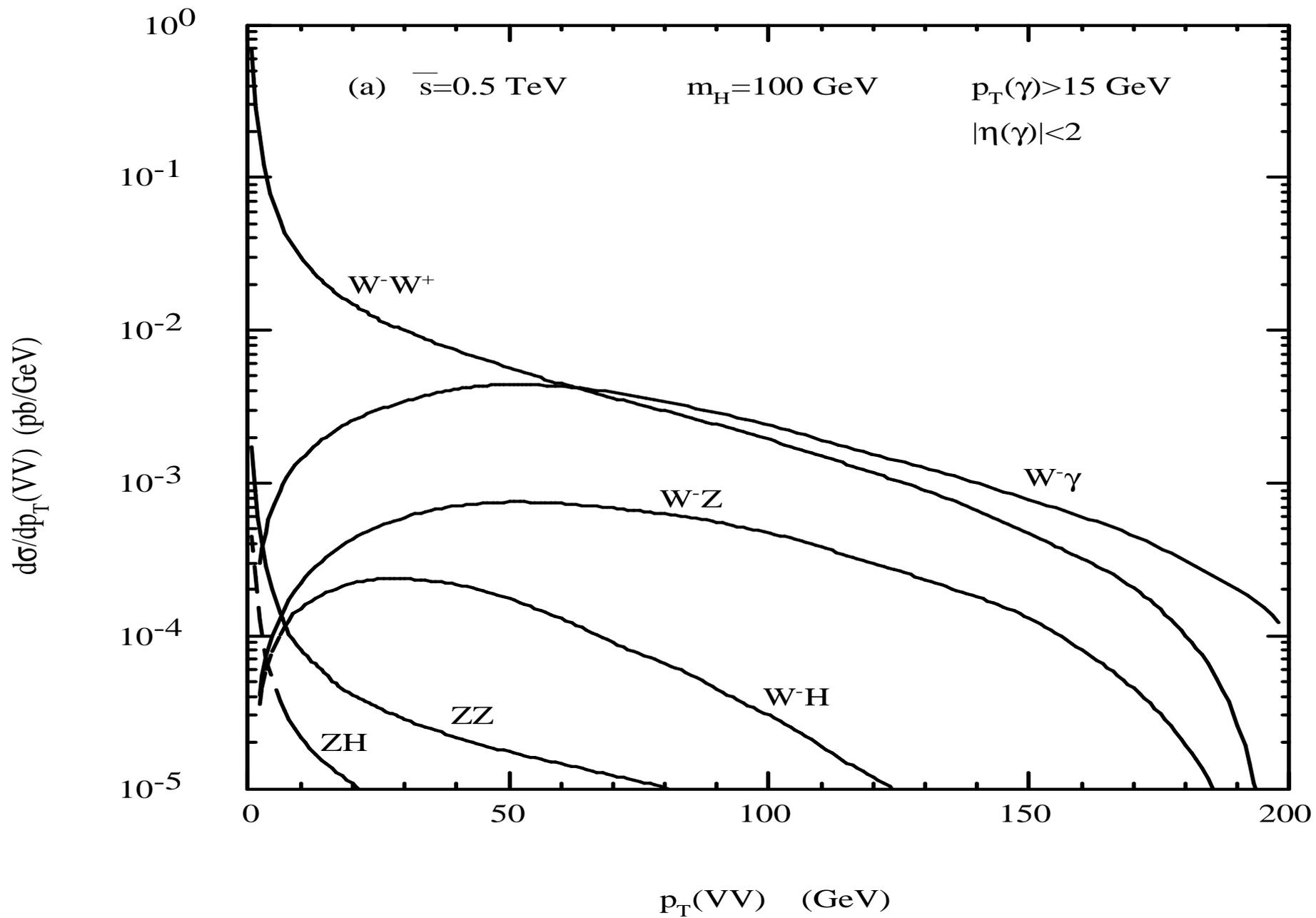
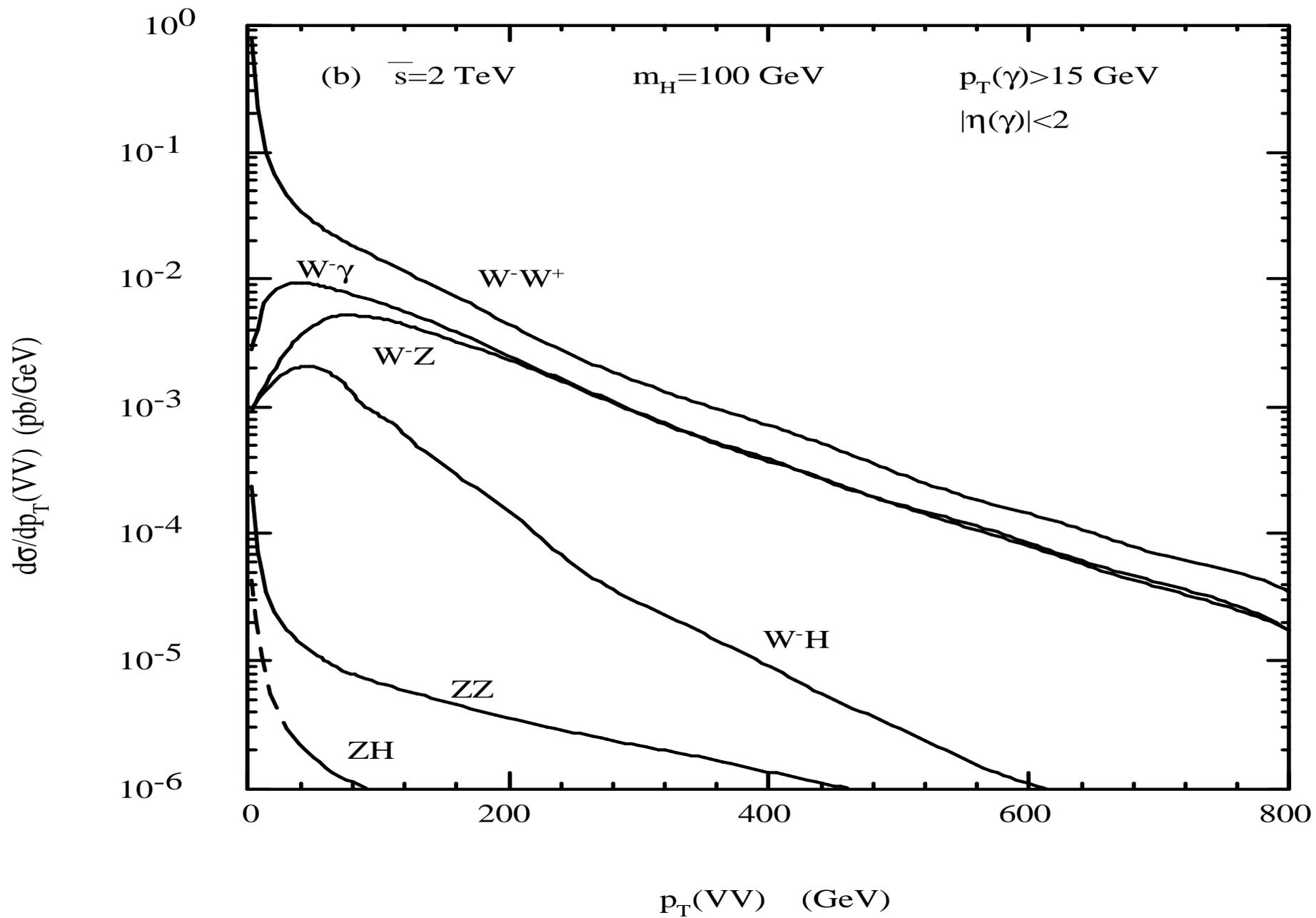
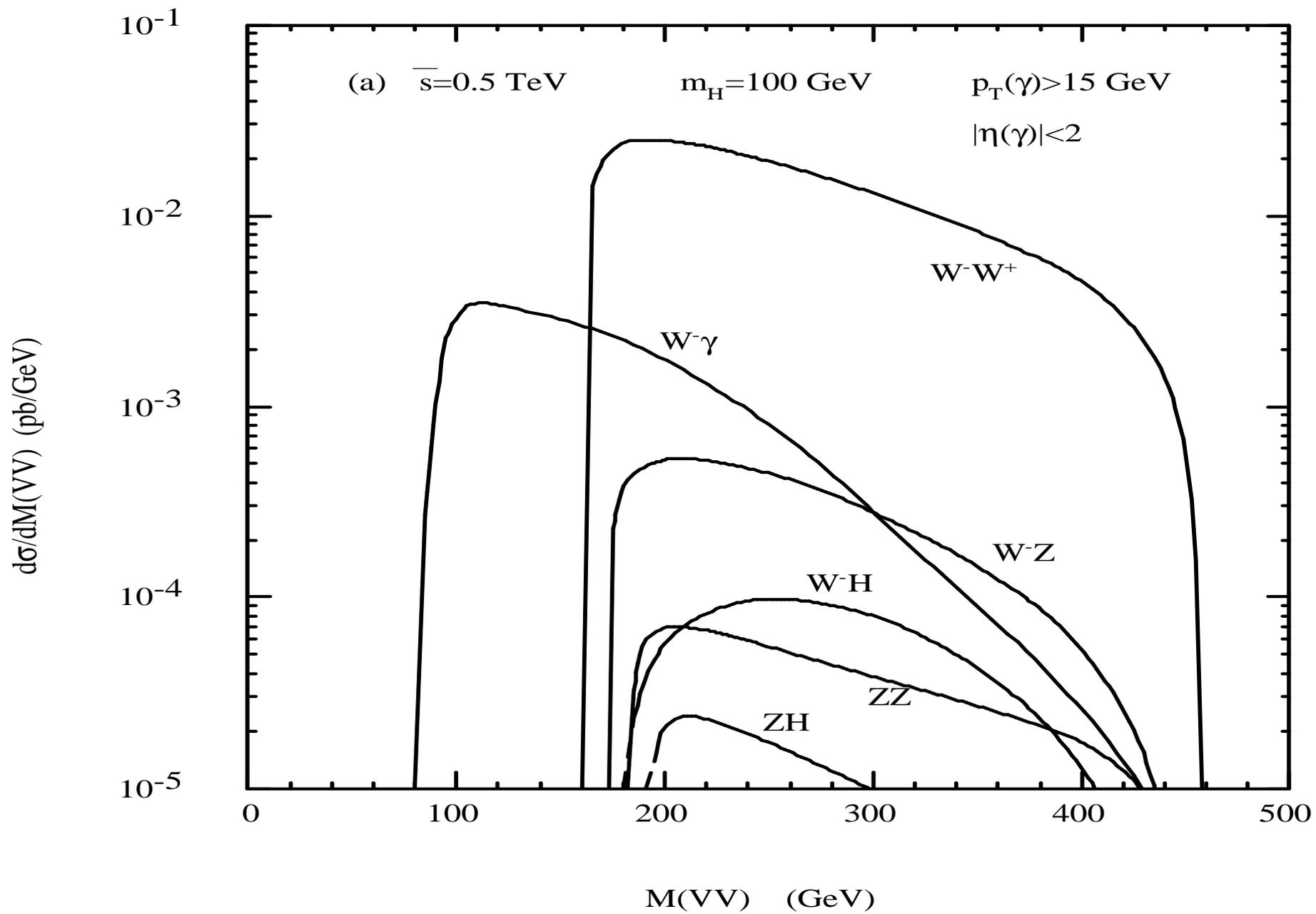
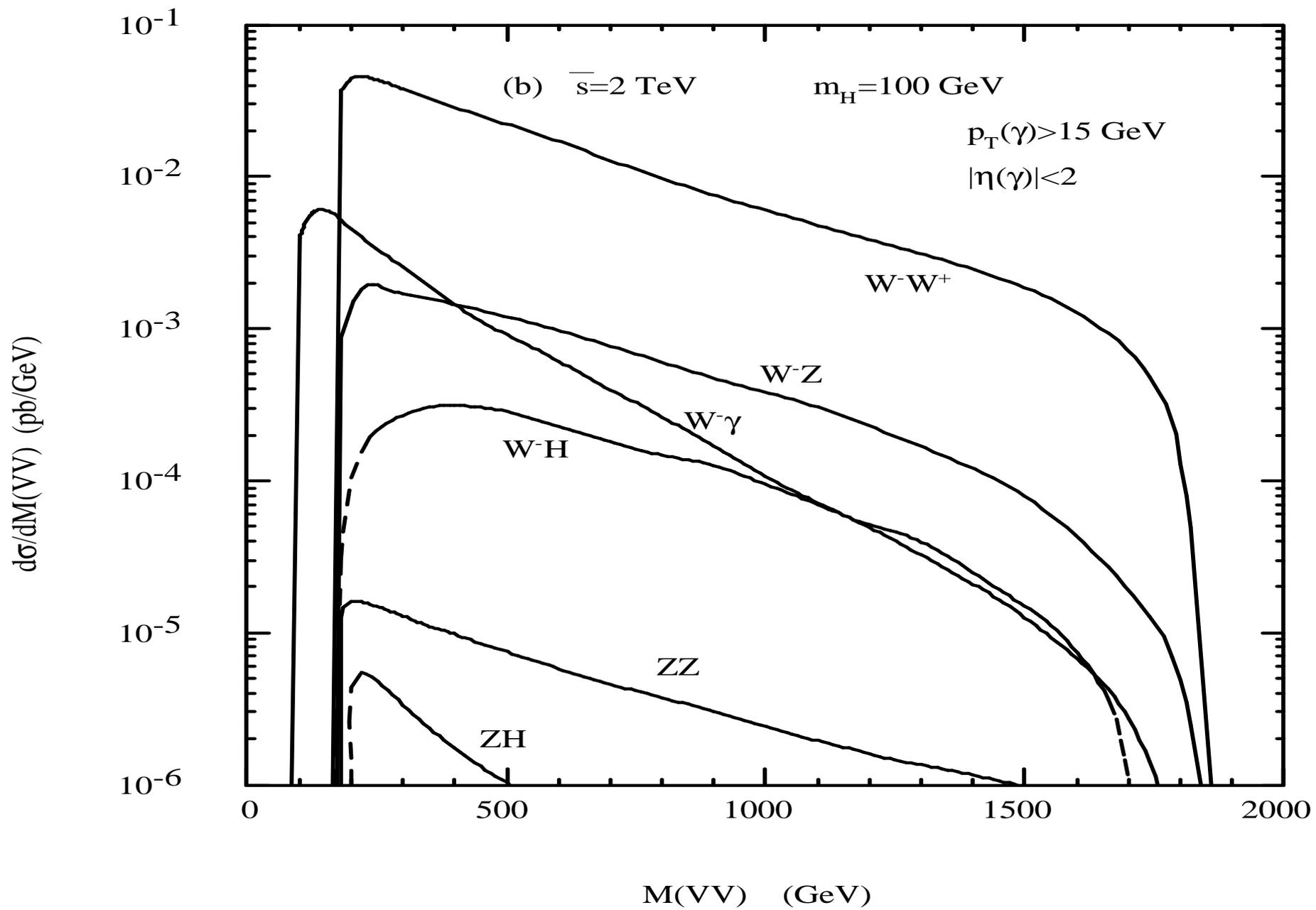
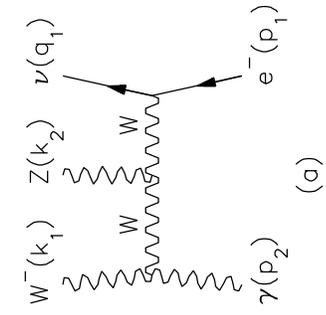
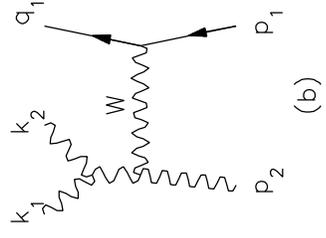
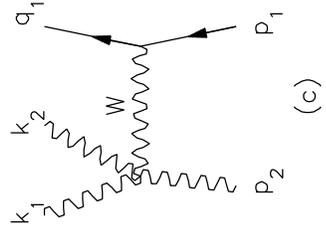
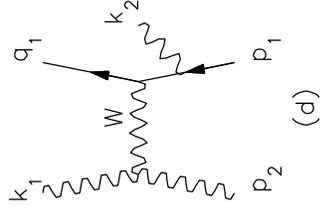
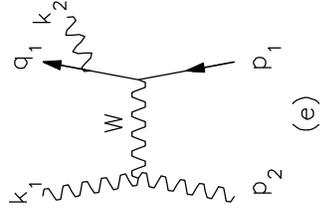
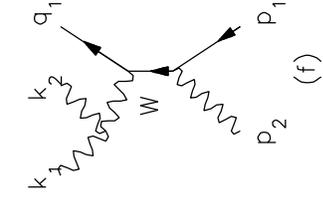
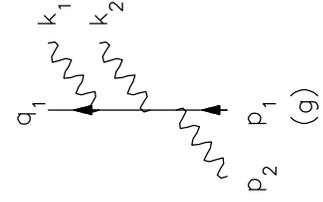
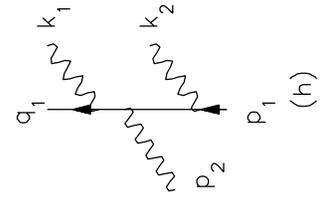
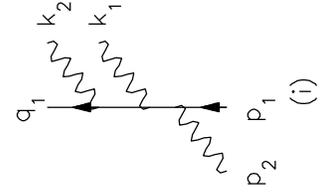
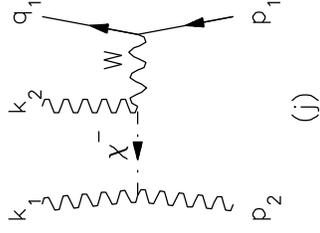
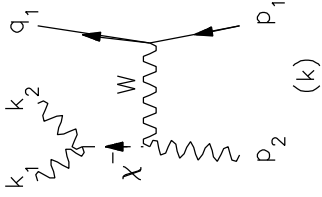
\end{document}